\documentclass{aa}

\newcommand{\ee} {\end{equation}}

\newcommand{\bc}{\begin{center}}
\newcommand{\ec}{\end{center}}
\def\ltsima{$\; \buildrel < \over \sim \;$}
\def\lsim{\lower.5ex\hbox{\ltsima}}
\def\loe{\lower.5ex\hbox{\ltsima}}
\def\gtsima{$\; \buildrel > \over \sim \;$}
\def\gsim{\lower.5ex\hbox{\gtsima}}
\def\goe{\lower.5ex\hbox{\gtsima}}
\newcommand{\be}{\begin{equation}}
\newcommand{\en}{\end{equation}}
\def\ergs{\rm \ erg \, s^{-1}}

\def\cmdue {\rm \ cm^{-2}}

\input psfig.tex

\begin{document}

%\thesaurus{ }

\title{A BeppoSAX view of transient black hole candidates in quiescence}

\titlerunning{A BeppoSAX view of BHCs in quiescence}
\authorrunning{Campana, Parmar \& Stella}
\author{S. Campana\inst{1} \and A.N. Parmar\inst{2} 
\and L. Stella\inst{3} }

\institute{Osservatorio Astronomico di Brera, Via E. Bianchi 46,
I--23807  Merate (Lecco), Italy
\and
Astrophysics Division, Space Science Department of ESA,
ESTEC, P.O. Box 299, 2200 AG Noordwijk, The Netherlands
\and
Osservatorio Astronomico di Roma, Via Frascati 33,
I--00040 Monteporzio Catone (Roma), Italy
}

\date{Received  / Accepted }
\offprints{S. Campana: campana@mera\-te.mi.astro.it}

\markboth{Campana et al.; BeppoSAX view of quiescent BHCs}
         {Campana et al.; BeppoSAX view of quiescent BHCs}

\abstract
{We report on BeppoSAX observations of five transient black hole candidates 
during their quiescent phase. We confirm these sources are X--ray faint,
improving on and complementing existing upper limits. We derive 1--10 keV
upper limits for GRO J0422+32, GRS 1009--45, 4U 1630--47 and XTE J1748--288, which 
range from $2\times 10^{32}\ergs$ to $5\times 10^{34}\ergs$. We positively
detect GS 2023+338. Its X--ray spectrum can be fit by a power law (photon 
index $\Gamma=1.9$) or a thermal bremsstrahlung (with $k\,T_{\rm br}= 7$ keV), 
converting in both cases to a 1--10 keV unabsorbed luminosity of $\sim 10^{33}\ergs$. 
These values are comparable to the ones derived during an ASCA observation 
in 1994, indicating that the source remains stable over a $\sim 5$ yr baseline.
}
\maketitle


\section{Introduction}

Transient black hole binaries are characterised by bright outbursts 
lasting up to several months and recurring every 1--100 yr (with a clear 
preference for long recurrence times). In between these outbursts, sources are 
quiescent with very low X--ray luminosities. At the present time only five 
transient Black Hole Candidates (BHCs) have been detected in quiescence 
and upper limits exist for six other systems (Asai et al. 1998; 
Menou, Narayan \& Lasota 1999a; Campana \& Stella 2000; Garcia et al. 2001).

A~0620--00 was the first BHC to be detected in quiescence. Despite the small 
number of photons collected in the ROSAT observation the spectrum was 
determined to be soft. A 0.5--10 keV luminosity of $L_{X}\sim10^{31}\ergs$ 
(for a distance of $d=1.2$ kpc) was obtained by extrapolating the ROSAT data 
(McClintock et al. 1995). Chandra detected the source at a level 
$\sim 10$ times lower (Garcia et al. 2001).
GS 2023+338 (V~404 Cyg) was  
observed with ASCA at a 0.5--10 keV luminosity of $L_{X}\sim 2\times 
10^{33}\ergs$ ($d=3.5$ kpc; Narayan, Barret \& McClintock 1997). The spectrum 
is well fit by a power law with photon index $\Gamma\sim 2.1^{+0.5}_{-0.3}$ 
or by a thermal bremsstrahlung with an equivalent temperature of $k\,T_{\rm br}=
4.6^{+3.6}_{-1.5}$ keV. Analysing the same dataset Asai et al. (1998) obtained 
$\Gamma=1.7^{+0.3}_{-0.2}$. Finally, GRO J1655--40 was detected again 
by ASCA at $L_{X} \sim 2\times 10^{32}\ergs$ ($d=3.2$ kpc; Hameury et al. 
1997). The spectrum could be described by a power law model with a quoted index 
$\Gamma\sim 1.5\pm0.6$ (Hameury et al. 1997). Asai et al. (1998) analysed the 
same dataset and found $\Gamma=0.7^{+2.1}_{-0.4}$. Chandra observed this source 
when it was a factor of $\sim 10$ (Garcia et al. 2001).

\begin{table*}
\caption{Summary of BeppoSAX observations.}
\begin{tabular}{ccccc}
Source       & LECS Exp. time & MECS Exp. time & PDS Exp. Time & Observation  \\
             & (s)            & (s)$^*$        & (s)           & number       \\
\hline
GRO J0422+32 &    19681       &    46274  (2)  & 21275         & 20535001 \\
GRS 1009--45 &    19773       &    40675  (2)  & 20689         & 20607001 \\
4U  1630--47 &    13357       &    32944  (3)  & 22426         & 20315001 \\
GS  2023+338 &     --         &    21384  (3)  & 10098         & 20303001 \\   
XTE J1748--288$^+$ & 19174    &    47884  (2)  & 34787         & 20130002 \\
\hline 
\end{tabular}

\noindent $^*$ in parenthesis the number of MECS active units.

\noindent $^+$ source at $5'$ off-axis.

\end{table*}

Since the discovery of their low X--ray quiescent luminosity, the study of BHCs 
in quiescence has attracted attention because this emission might be able 
to distinguish them from quiescent neutron star systems (for a review see 
Campana et al. 1998a). 
At the very beginning, Tanaka \& Shibazaki (1996) argued that BHCs and 
neutron stars systems at low luminosities show similar soft spectra, well 
approximated by a single temperature black body 
($k\,T_{\rm bb}\sim 0.2-0.3$ keV). 
This was based (and biased) on ROSAT results and only on a single BHC 
(A~0620--00). ASCA and BeppoSAX observations led to the discovery of hard 
tails in neutron star 
systems (Asai et al. 1996, 1998; Campana et al. 1998b, 2000) as well as 
the detection of non-thermal spectra in GS 2023+338 and GRO J1655--40,
challenging this picture (however, GS 2023+338 and GRO J1655--40 have  
orbital periods longer than the great majority of low mass transients and, 
therefore, likely experience a higher time-averaged mass exchange rate).
In a recent paper, Garcia et al. (2001) reported on deep Chandra observations
of black hole transients in quiescence, detecting two new sources with short
orbital period (GRO J0422+32 and GS 2000+25). Now three short orbital period BHCs 
have been detected all with 0.5--10 keV luminosities in the 
$2-8\times 10^{30}\ergs$ range. These data confirm that black hole transients 
in quiescence are much less luminous than the corresponding neutron star 
transient systems {\it in X--rays}.

Advection dominated accretion flow (ADAF) models have become popular in 
explaining the low luminosity of quiescent BHCs as well as their spectral 
energy distribution (Narayan et al. 1996; Narayan, Garcia \& McClintock 
1997; Menou et al. 1999a, 1999b). In these models a large fraction of the 
gravitational energy is advected into the black hole, therefore lowering 
considerably the radiative efficiency of the accretion process. 
Observations show that the ratio of minimum X--ray luminosity in quiescence to 
maximum X--ray luminosity in outburst is significantly smaller (a factor of
about 100) in black hole transients than in and neutron star transients 
(Narayan et al. 1997; Garcia et al. 1998; Garcia et al. 2001). 
Campana \& Stella (2000) noted that in the latest ADAF models the optical/UV 
luminosity derives from synchrotron radiation produced by the ADAF itself and 
therefore must be included in the luminosity budget.
Ascribing the bulk of the residual optical/UV flux to the ADAF removes much 
of the difference in luminosity swing between black hole and neutron star 
transients, weakening one of the main drives of ADAF models.
Possible ways out are that a large fraction of the accreting matter is 
stopped/evaporated at the outer disk boundary with only a very small fraction 
leaking towards the compact object and/or ADAFs do not contribute to the 
optical luminosity as in the older ADAF models (Campana \& Stella 2000).

In this paper we present a comprehensive view of black hole transients in 
quiescence observed with the Italian/Dutch satellite BeppoSAX complementing
and improving current upper limits and existing spectra.
In Section 2 we describe the data set and analysis. In Section 
3 we discuss the results.

\section{Observations and data reduction} 

We present the results of the observations of black hole transients in 
quiescence carried out with the BeppoSAX satellite (Boella et al. 1997a).
We analysed the data from the Low Energy Concentrator Spectrometer 
(LECS; 0.1--10 keV, Parmar et al. 1997a) and the Medium Energy Concentrator 
Spectrometer (MECS; 1.3--10 keV, Boella et al. 1997b).  
As usual, LECS data were collected only during satellite night-time leading 
to shorter exposure times. 
Upper limits in the hard X--ray band were obtained with the
Phoswich Detector System (PDS, 15--300 keV; Frontera et al. 1997). 
The PDS collimators rocked on and off the target in order to monitor the 
background. This also resulted in a shorter exposure time than the MECS.

BeppoSAX observed four BHCs in quiescence: GRO J0422+32, GRS 1009--45, 4U  
1630--47 and GS 2023+338. XTE J1748--288 was serendipitously observed during a 
pointing toward the Galactic Center region. For a summary of the observations 
see Table 1. 

\begin{figure*}[!htb]
\psfig{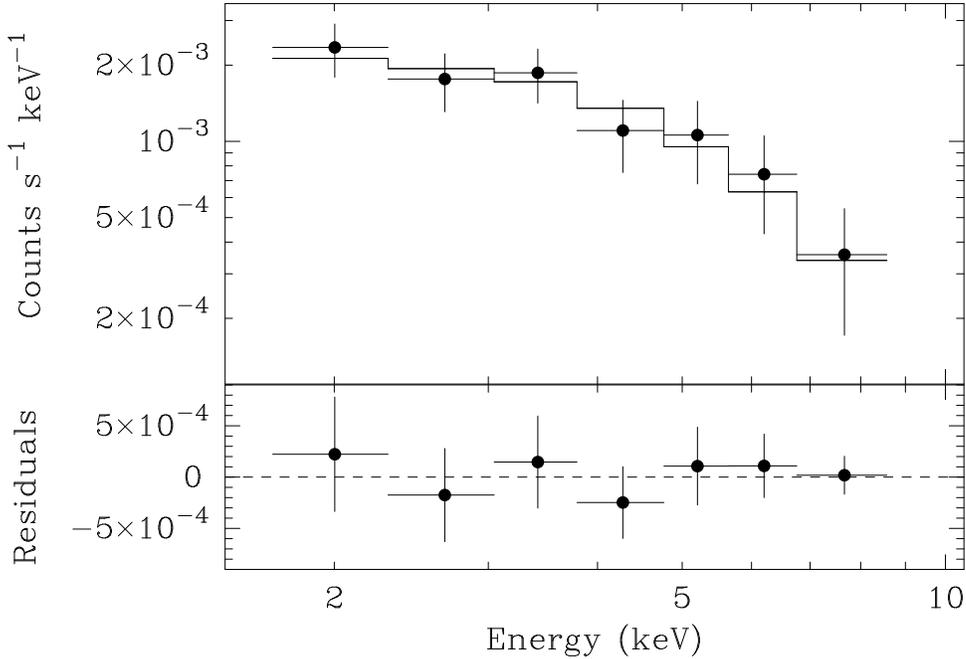}
\caption{X--ray spectrum of GS 2023+338 in quiescence.
The spectrum is fit with an absorbed power law model.
In the lower panel the ratio between the data and the model
is presented.}
\label{figdue}
\end{figure*}

The BeppoSAX images were searched for sources at the position of the 
BHCs' optical counterpart. Only one out of five BHCs was detected (GS 2023+338) 
confirming the elusiveness of this kind of source in quiescence. 

Upper limits for imaging instruments were determined counting the number of 
events within a specified box ($4'$ radius), correcting those counts for 
vignetting, exposure and point spread function and comparing them with the 
number of events in an outer background box free of sources. 
In the case of the PDS collimators source upper limits were derived by 
comparing the number of events collected on and off the the source.

GRO J0422+32 was observed by BeppoSAX on September 2, 1998, about 2,200 days
after the previous outburst. The source was not detected and we place a 
$3\,\sigma$ 
upper limit of $2\times 10^{-3}$ c s$^{-1}$ on the MECS count rate, which 
translates 
to a flux limit of $1.5\times 10^{-13}\ergs\cmdue$ (2--10 keV), assuming a power 
law spectrum with photon index $\Gamma=2.1$ (the same as observed in GS 2023+338 
in quiescence; Narayan et al. 1997) and the galactic column density in the 
direction of the source $N_H=1.7\times 10^{21}\cmdue$. For a distance of 3.6 kpc 
(Shrader et al. 1994) we obtain an upper limit on the 2--10 keV luminosity of 
$2\times 10^{32}\ergs$. 
This is to be compared with the value derived by Garcia et al. (1998) of 
$4\times 10^{31}\ergs$ (0.5--10 keV) which was obtained by extrapolating over 
a much larger energy range the result of a deep ROSAT-HRI pointing.
The BeppoSAX PDS data give an upper limit of $\sim 6\times 10^{-12}\ergs\cmdue$ 
in the 15--50 keV energy band, which translates to $\sim 9\times 
10^{33}\ergs$.

4U 1630--47 was observed on March 26, 1997, about one year after the 1996 outburst
(this source in contrast with other transient BHCs shows recurrent outbursts, 
Parmar, Angelini \& White 1995). The $3\,\sigma$ 
upper limit on the MECS count rate amounts to $1.2\times 10^{-3}$ c s$^{-1}$.
This translates into an upper limit on the 2--10 keV unabsorbed luminosity
of $2\times 10^{33}\ergs$ for a distance of 10 kpc, a power 
law with $\Gamma=2.1$ and $N_H=8\times 10^{22}\cmdue$ (Oosterbroek et al. 1998). 
Even if this upper limit is in the range of those previously obtained with the
ROSAT-HRI ($8\times 10^{33}\ergs$ 0.2--2.4 keV) and ROSAT-PSPC ($7\times 
10^{32}\ergs$ 0.2--2.4 keV; Parmar et al. 1997b), the more extended energy range 
of the MECS makes the BeppoSAX limit less affected by interstellar absorption. 
At high energies, the PDS limit on the 15-50 keV flux was of $\sim 9\times
10^{-11}\ergs\cmdue$ (corresponding to $\sim 1\times 10^{36}\ergs$). This 
relatively high value is due to the presence of a number of sources in the PDS 
field of view, which are clearly seen in the MECS.

GRS 1009--45 was observed on December 3, 1998, about 1,900 days after the
previous outburst. We obtain a $3\,\sigma$ upper limit of $1.6\times 10^{-3}$ 
c s$^{-1}$ on the MECS count rate. This translates into 
an upper limit of $1\times 10^{33}\ergs$ on the 2--10 keV unabsorbed 
luminosity for a distance of 3 kpc (Della Valle et al. 1997), 
a power law with $\Gamma=2.1$ and $N_H=10^{21}\cmdue$. 
The 15--50 keV upper limit from the PDS amounts to $\sim 7\times 10^{33}\ergs$. 
No previous upper limits exist on the 2--10 keV and 15--50 keV luminosities.

XTE J1748--288 was observed on April 5, 1997, about 400 d before its discovery.
We obtain a $3\,\sigma$ upper limit on the MECS count rate of $6\times 10^{-3}$ 
c s$^{-1}$. 
Despite the relatively long exposure, this fairly high value is due 
to the source position right on the MECS ``strongback'' (Boella et al. 1997b). 
Using the appropriate response matrix we derive a flux of $<6\times 
10^{-12}\ergs\cmdue$ (i.e. $<5\times 10^{34}\ergs$ at 8.5 kpc). No upper 
limits in the 15--50 keV energy band can be set due to the presence of strong 
sources in the field.

In passing we note that GX 339-4 has been observed and detected by BeppoSAX
at a level of $\sim 6\times 10^{33}\ergs$ (Kong et al. 2000). This luminosity level
is high when compared with the other BHCs, suggesting that the observed 
`off'-state of GX 339-4 may not correspond to the quiescent state of a transient 
system (note also that GX 339-4 is usually considered a persistent source).
Another interesting system monitored by BeppoSAX is CI Cam (XTE J0421+560).
This source was observed three times in quiescence (156, 541 and 690 d after 
the latest outburst, respectively) with very different outcomes:
$i$) a very soft spectrum ($\Gamma\sim 4$) with a low column density ($N_H\sim 
10^{21}\cmdue$) and a 1--10 keV luminosity of $2\times 10^{32}\ergs$ (at 2 kpc); 
$ii$) a hard spectrum ($\Gamma\sim 2$) with a high column density 
($N_H\sim 4\times 10^{23}\cmdue$) and a luminosity of $4\times 10^{33}\ergs$; 
$iii$) undetected with an upper limit of $4\times 10^{32}\ergs$ (Parmar et al. 2000).
However, the nature of the compact object of CI Cam is presently unclear.

\subsection{Spectral analysis: GS 2023+338}

The source spectrum of GS 2023+338 was extracted from the MECS data within 
a radius of $4'$ centered on the source position (LECS data were unavailable). 
We collected 433 photons within the full energy range. Background subtraction was 
performed using the standard background files. We used the publicly available 
calibration files at 2000 January and XSPEC 11.0. We rebinned the MECS spectrum 
in order to have at least 50 photons per spectral bin, resulting in seven bins. 
The lack of the LECS data forced us to adopt a value for the absorbing column 
density. Following Wagner et al. (1994) and Narayan et al. (1997) we used
a value of $N_H=10^{22}\cmdue$. With this value of the column density 
a power law model provides an adequate description of the data with a photon 
index $\Gamma=1.9^{+0.6}_{-0.5}$ (errors at 90\% confidence level for one 
parameter of interest) with a $\chi^2_{\rm red}=0.2$ (for 5 d.o.f.). A 
bremsstrahlung model provides also a good fit with 
$k\,T_{\rm br}=6.9^{+31.5}_{-3.7}$ 
keV and $\chi^2_{\rm red}=0.3$. A black body model instead fails to 
successfully describe the spectra with $\chi^2_{\rm red}=1.4$ ($k\,T_{\rm bb}=
1.0^{+0.3}_{-0.2}$ keV). This last model can 
be reconciled with the data only by assuming a very low column density (formally 
null) and $k\,T_{\rm bb}=1.1^{+0.1}_{-1.1}$ keV ($\chi^2_{\rm red}=1.1$). 
The unabsorbed 1--10 keV flux 
as derived with the power law model amounts to $7.1\times 10^{-13}\ergs\cmdue$.
This is to be compared with the $8.2\times 10^{-13}\ergs\cmdue$ value derived 
from the ASCA observation (Narayan et al. 1997). The 1--10 keV unabsorbed 
luminosity is $L_X=1.2\times 10^{33}\ergs$ (at 3.5 kpc).
The PDS provides an upper limit to the 15--50 keV luminosity of $\sim 3\times 
10^{34}\ergs$.

\section{Conclusions}

We report on the BeppoSAX view of transient BHCs in quiescence. We further
confirm that BHCs in quiescence are X--ray faint with luminosities below
$\sim 10^{32}-10^{33}\ergs$ (at which neutron star transients are usually 
detected) and reporting for the first time upper limits in the 2--10 keV 
energy band for four sources GRO J0422+32, GRS 1009--45, 4U 1630--47 and 
XTE J1748--288. These limits improve and confirm existing extrapolations 
from lower energy bands (mainly ROSAT, 0.1--2.4 keV). 
Moreover, we quote the upper limits in the hard energy band 15--50 keV 
derived from the PDS data. These hard X--ray luminosity limits might be useful 
to constrain ADAF models, for which high energy tails are expected (current 
models easily satisfy these constraints, e.g. Menou, Narayan \& Lasota 1999a). 
After the failure of ASTRO-E, these limits will remain unrivaled for years to come.

One exception to the above cases, that was confirmed also by BeppoSAX 
observation, is GS 2023+338 revealed at a level of $\sim 10^{33}\ergs$ 
(1--10 keV). The GS 2023+338 luminosity derived is at a level and with a 
spectrum similar to those of the ASCA observation five years before (Asai et al. 
1998; Hameury et al. 1997). This indicates that, within the uncertainties, 
the spectrum and the flux level of GS 2023+338 in quiescence remains fairly
constant.

\begin{acknowledgements}
We thank T. Mineo for making available to us off-axis MECS response matrices.
This research has made use of SAXDAS linearised and cleaned event
files (Rev.2.0) produced at the BeppoSAX Science Data Center.
This work was partially supported through ASI grants.
\end{acknowledgements}

\end{document}